\documentclass{aastex}
\usepackage{graphics}
\usepackage{subfigure}
\usepackage{lscape}
\usepackage{emulateapj5}

\begin{document}

\title{M--type Vega--like Stars}

\author{Inseok Song\altaffilmark{1}, A. Weinberger\altaffilmark{2}, E. E. Becklin, B. Zuckerman,
C.~Chen}

\affil{Dept. of Physics and Astronomy -- University of California, Los~Angeles}

\affil{8371 Math Science Bldg. -- Box~951562 }

\affil{Los~Angeles, CA 90095--1562, USA}

\email{song@astro.ucla.edu, weinberger@dtm.ciw.edu, ben@astro.ucla.edu, becklin@astro.ucla.edu,
cchen@astro.ucla.edu}

\altaffiltext{1}{UCLA Center for Astrobiology post--doctoral fellow}
\altaffiltext{2}{Current address: Dept. of Terrestrial Magnetism, Carnegie Institute of Washington,
5241 Broad Branch Rd. Washington, DC 20015}

\begin{abstract}
We carried out a search for M--type Vega--like stars by correlating
the \emph{IRAS Faint Source Catalog} with \emph{Hipparcos} selected
M--type stars. Three stars with apparent \emph{IRAS} 25~$\mu$m
excess emission are shown instead to be non--IR--excess stars from
ground--based 11.7 and 17.9~$\mu$m photometry. Two stars previously
suggested to have Vega--like mid--IR excess are also shown to be non--excess
stars. These results imply that other suggested mid--IR excess stars
in the literature may also be false excess stars. Detection threshold
bias is apparently responsible for these bogus IR excesses. Sixty
micron excess emission from a previously known M--type Vega--like
star (GJ~803) is identified again. 
\end{abstract}


\section{Introduction}

During the past fifteen years, about two dozen papers have been published
that describe searches for stars with excess infrared (IR) emission
(for recent reviews on these {}``Vega--like{}'' stars, see \citealt{LBA99}
and \citealt{Zuckerman}). These searches employed different techniques
for cross correlating IR and stellar sources with no consistent definition
of what defines an IR excess (see \citealt{myPhD} for a summary).
To date, about 400 Vega--like stars have been identified.  The Vega phenomenon
overlaps with some very important solar system formation epochs; gas giant
planet formation at $\lesssim10$~Myr and terrestrial planet formation at
$\sim100$~Myr. Thus, knowing stellar ages of Vega--like stars is 
essential to studying extra solar planetary system formation in detail.
Consequently, there have been many efforts to estimate
the ages of Vega--like stars \citep[see][ and references therein]
{myPhD,Spangler}.
Stellar ages can be fairly accurately and relatively easily determined
for late--type stars (e.g., \citealt{lateVegas}), however, 
ages of early--type stars are less reliable (\citealt{AVegas}). Most
of the currently known Vega--like stars are early--type because more
luminous stars produce larger IR fluxes for stars with the same quantity of
dust and the surveys conduced by IRAS were flux limited. Thus, it
is desirable to increase the known number of late--type Vega--like
stars for more precise age estimates of such stars.  Additional identification 
of late--type Vega--like stars is very useful to statistically strengthen 
studies such as planetary formation in different environments and dust 
lifetime as a function of stellar mass and luminosity (see \citeauthor
{2Vegas}'s (2000) suggestion of dichotomy of Vega--like stars for example). 
Also, late--type Vega--like stars are excellent laboratories for studying 
the early evolution of our solar system.

Despite the great number of M-type stars compared to earlier types, only
two with IR excesses, GJ~803 \citep{Tsikoudi} and Hen~3-600
\citep{delaReza,RayJay} have been identified to date.  Almost all
previous studies (exceptions include \citealt{AP} and \citealt{Od})
searched only for far-infrared (60~$\mu$m) excess and were limited by
the IRAS $\sim$90\% completeness sensitivity of 280~mJy in that band.  IRAS
sensitivity at 25~$\mu$m was 210~mJy, and an M0 star located 10~pc away
with 200~K dust grains absorbing and emitting like blackbodies such that
L$_{IR}$/L$_{star}$=1.5 $\times 10^{-3}$, would have been detected by
IRAS at 25~$\mu$m (F$_{star+disk}\simeq250$~mJy) but not at
60~$\mu$m (F$_{star+disk}\simeq100$~mJy).

A more complete survey for dust must be able to detect the stellar photosphere
at high precision in order to evaluate excess emission above that level.
One can do a fairly thorough disk excess assessment for stars whose
photospheres were detected by IRAS. However, around stars whose
photospheres were too faint to be detected, only unusually large disk
excesses could be detected.  At 60
$\mu$m, IRAS could detect the photosphere of an A0 star out to $\sim$20~pc, but
an M0 only out to $\sim$2~pc. The situation is somewhat better at 25 $\mu$m,
where IRAS could detect the photosphere of an A0 star out to $\sim$50~pc and an
M0 photosphere out to $\sim$5~pc.  By searching the catalog only at 60 $\mu$m,
previous surveys have not probed large regions of phase space where excess
may exist around late-type stars.

In this study, we concentrated mainly on M--type stars and attempted
to perform the most thorough search for M--type Vega--like stars to
date, especially at 25~$\mu$m, based on the \emph{IRAS} FSC \citep{FSC} 
and \emph{Hipparcos} catalog \citep{Hipparcos}. 
As a check on recently reported infrared excess stars, however, we also 
report on the F--type star HD~2381.

\section{Search \label{search}}

Based on the \emph{Hipparcos} catalog, we selected $\sim530$ nearby 
(\( <25 \)~pc) stars with  $(B-V)+\sigma_{(B-V)} >1.40$, where 
$\sigma_{(B-V)}$ is uncertainty of $(B-V)$.  The \emph{Hipparcos} catalog
contains almost all early M--type (M0--2) stars within 10~pc. Many
IR sources in the FSC with optical stellar identifications are giant
stars \citep{ZKL,Od}. Therefore, one needs luminosity class information
to identify Vega--like stars. Following \citet{Silverstone}, we used
a constraint on the absolute visual magnitude 
(\( M_{V}>7.5\times \left( B-V\right) -5.0 \))
to ensure that a candidate is not a giant star whose IR--excess mechanism
may be different from that of a dwarf. 
Six stars from our initial sample do not meet the absolute visual magnitude
cut and they are HIP~21421, 50798, 66212, 66906, 75187, and 82099. HIP~21421
and HIP~66212 are K--type giants and the other four stars appear to be main
sequence stars with large uncertainties in $(B-V)$. Among the four rejected main
sequence stars, HIP~75187 is the only one detected by {\em IRAS} (only
at 12~$\mu$m) and the measurement agrees with the flux density expected
from its photosphere alone.
Then, our sample
stars were cross correlated with FSC sources with a maximum allowed
offset of \( 30'' \) between \emph{Hipparcos} and \emph{IRAS} source
positions (both at epoch 1983.5 and equinox 1950). Only 152 stars
from our initial sample have IR counterparts; among them, 96 stars
were detected only at 12~$\mu$m and 55 stars were detected
at both 12 and 25~$\mu$m. GJ~803 (AU~Mic) was detected at
12 and 60~$\mu$m and was the only dwarf M--type star that has
been detected at the \emph{IRAS} 60~$\mu$m band. No objects
were detected at 100~$\mu$m. A previously known M--type IR
excess star, Hen~3--600 (TWA~3), is not identified in our search
because it is not bright enough to be included in the \emph{Hipparcos}
catalog. 

Positions given in the published \emph{IRAS} catalog are weighted
means of 12, 25, 60, and 100~$\mu$m source positions based
on their signal-to-noise ratios. Sometimes, 25 and 60~$\mu$m
sources are background objects far away from the stellar 12~$\mu$m
sources. Thus, in IR--excess surveys, it is mandatory to check each
band's source position and to confirm that source positions in each
band are coincident. We checked the offsets among 12 and 25~$\mu$m positions 
(60~$\mu$m position also for GJ~803) by using the {}``LONG FSC{}'' from
\emph{The Infrared Processing and Analysis Center (IPAC)} at Caltech
and found that only one object (GJ~433) shows a substantial (36$''$)
offset between the 12 and 25~\( \mu m \) \emph{IRAS} sources. For comparison, 
a median positional uncertainty of IRAS FSC sources in the cross scan direction is
20$''$. Offsets
between the 12 and 25~$\mu$m source positions for the other
stars are negligible with respect to the \emph{IRAS} positional uncertainties. 

To identify IR excesses, we performed spectral energy distribution
(SED) fitting by using all known photometric data from the literature
(queried through \emph{SIMBAD}) including online \emph{2MASS} data.
For GJ~413.1 and GJ~433, JHK magnitudes (see Table~\ref{table})
were measured on 24 November, 2000 (UT) with the NASA IRTF telescope
at Mauna~Kea Observatory. Since the accuracy of the photospheric
flux estimation at 12~$\mu$m depends strongly on the availability
of near IR photometric data (i.e., JHK magnitudes), we have not carried
out a SED fit to stars with only 12~$\mu$m detections because
most of these stars lack near IR photometry. Stellar SEDs are different
from that of a blackbody. Opacity sources absorb light at wavelengths
with high opacities and re-radiate it at wavelengths with relatively low 
opacities. This results in a spectral energy distribution very different 
from that of a black-body. Therefore we used \textsc{PHEONIX}
NextGen synthetic stellar spectra \citep{NextGen} instead of a blackbody
SED. Among three SED fitting parameters (parallax, stellar radius,
and effective temperature), \emph{Hipparcos} parallax has been treated
as constant. Stellar radius and effective temperature were estimated
from the \emph{Hipparcos} \( (B-V) \) value by using the spectral
type versus colors/\( T_{eff} \)/radius relation of \citet{MKspec}. 

To quantify the strength of IR excesses, we defined \( r \), the
specific IR excess, (``specific excess'' hereafter) as

\begin{equation}
\label{r}
r\equiv \frac{F_{25\, \mu m}^{IRAS}-F_{25\, \mu m}^{est}}{F^{est}_{25\, \mu m}}
\end{equation}
where \( F_{25\, \mu m}^{IRAS} \) and \( F_{25\, \mu m}^{est} \)
are \emph{IRAS} FSC 25~$\mu$m flux and estimated photospheric
contribution at 25~$\mu$m, respectively. For GJ~803, we used
\( F_{60\, \mu m}^{IRAS} \) and \( F_{60\, \mu m}^{est} \) to calculate
its 60~$\mu$m specific excess (\( r_{60\, \mu m} \)). 

\begin{figure}
{\centering \resizebox*{1\columnwidth}{!}{\includegraphics{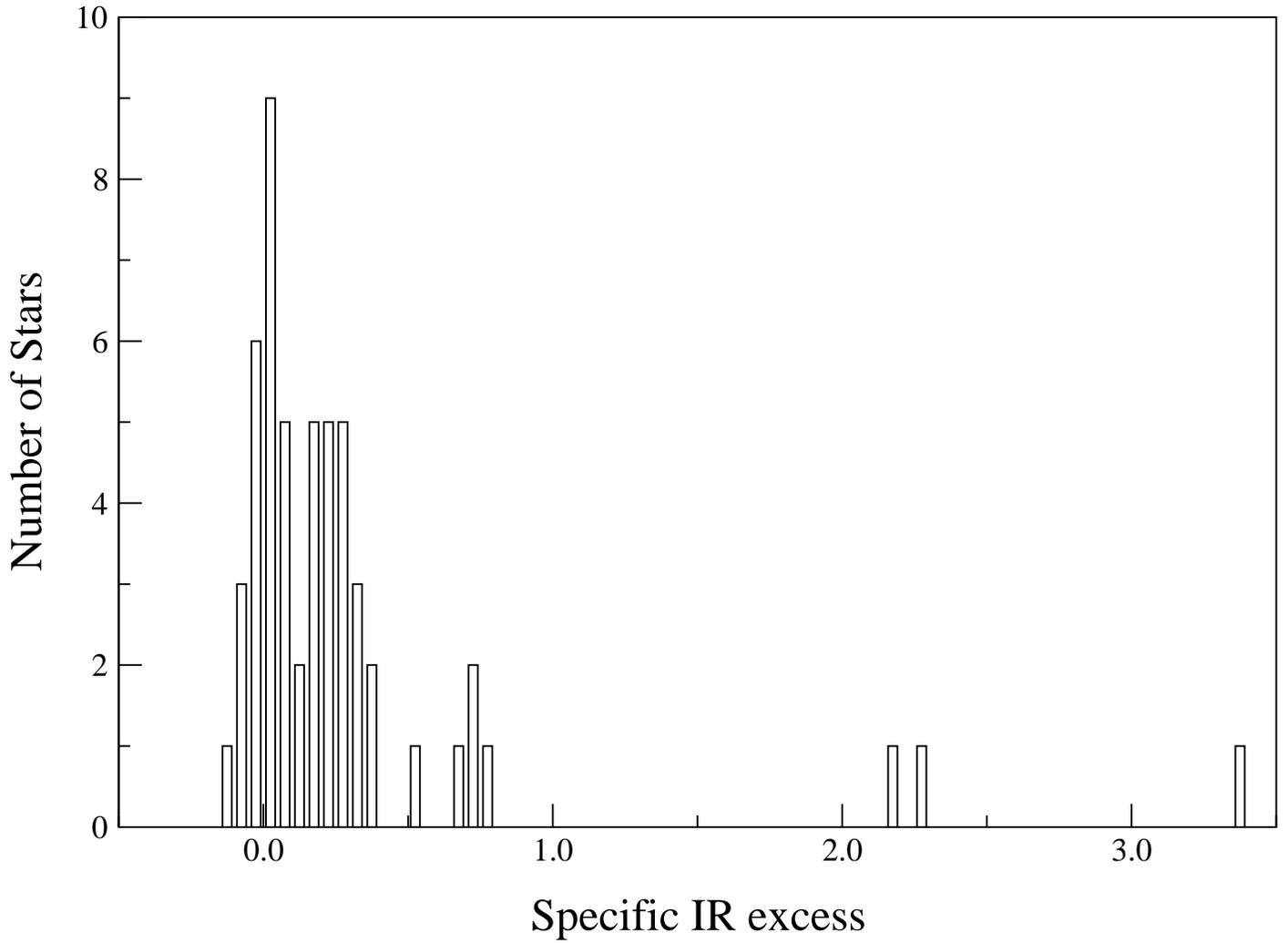}} \par}

\caption{Histogram of specific IR excess (\protect\( r_{25\, \mu m}\protect \))
for the 55 stars discussed in the text. The \protect\( r\protect \)-value
of GJ~803 (\protect\( r_{60\, \mu m}=7.6\protect \)) is outside
of the displayed range. \label{histogram}}
\end{figure}

As shown in Figure~\ref{histogram}, we found three stars (GJ~154,
413.1, and 433) with \( r>2.0 \) based on 25~$\mu$m fluxes
and a different star (GJ~803) with \( r=7.60 \) based on 60~$\mu$m
flux. Contrary to an expected median specific excess value of zero for
non--IR--excess stars, Figure~\ref{histogram} shows a median value
of \( \sim 0.1 \) which may be due to the 25~$\mu$m flux overestimation
as explained in the \emph{IRAS} Explanatory Supplement Version~2,
III--131. All IRAS flux density values in Table~1 and Figures~2--3 
are color corrected using Table~VI.C.6 of the \emph{IRAS} Explanatory 
Supplement Version~2. For 12 and 25~$\mu$m fluxes, stellar effective temperatures
were used to estimate color correction factors. However, for 60~$\mu$m fluxes, 
if any IR excess exist (e.g., GJ 803), then dust temperatures were used instead
of stellar effective temperatures.

\section{Ground--based mid--IR Photometry}

Mid--infrared imaging was performed with the facility instrument,
the Long Wavelength Spectrograph (LWS) \citep{Jones93}, on the 10~m
Keck~I telescope on UT 11 December 2000 and 4--5 February 2001. During
all three nights, the weather was photometric with low water vapor
optical depth. LWS uses a \( 128\times 128 \) pixel Boeing Si:As
detector, and has a plate scale of 0.08 arcsec/pixel, resulting in
a focal--plane field of view of \( 10.^{''}24\times 10.^{''}24 \).
Each object was measured in filters centered at 11.7~$\mu$m
(FWHM=1.0~$\mu$m) and 17.9~$\mu$m (FWHM=2.0~$\mu$m).
Images were obtained at four positions by chopping the secondary at 
2.5--5~Hz with a throw of $10''$ and nodding the telescope $10''$ after \( \sim 20 \)~s.
In basic data reduction, the images were double differenced to remove
the sky and telescope background, and bad pixels were corrected by
interpolation. Throughout the nights, including just before and after
each of the M--star measurements, bright infrared standard stars were
observed for photometric calibration. Standard star measurements over
the whole of each night were averaged and the standard deviation in
their photometry was used as an estimate of calibration uncertainty.
On 11 December, the uncertainty in the calibration was 5\% and 6\%
at 11.7 and 17.9~$\mu$m, respectively. On 4 and 5 February,
the uncertainties were 15\% at both wavelengths.

\begin{figure}
{\centering \begin{tabular}{cc}
\resizebox*{0.5\columnwidth}{!}{\includegraphics{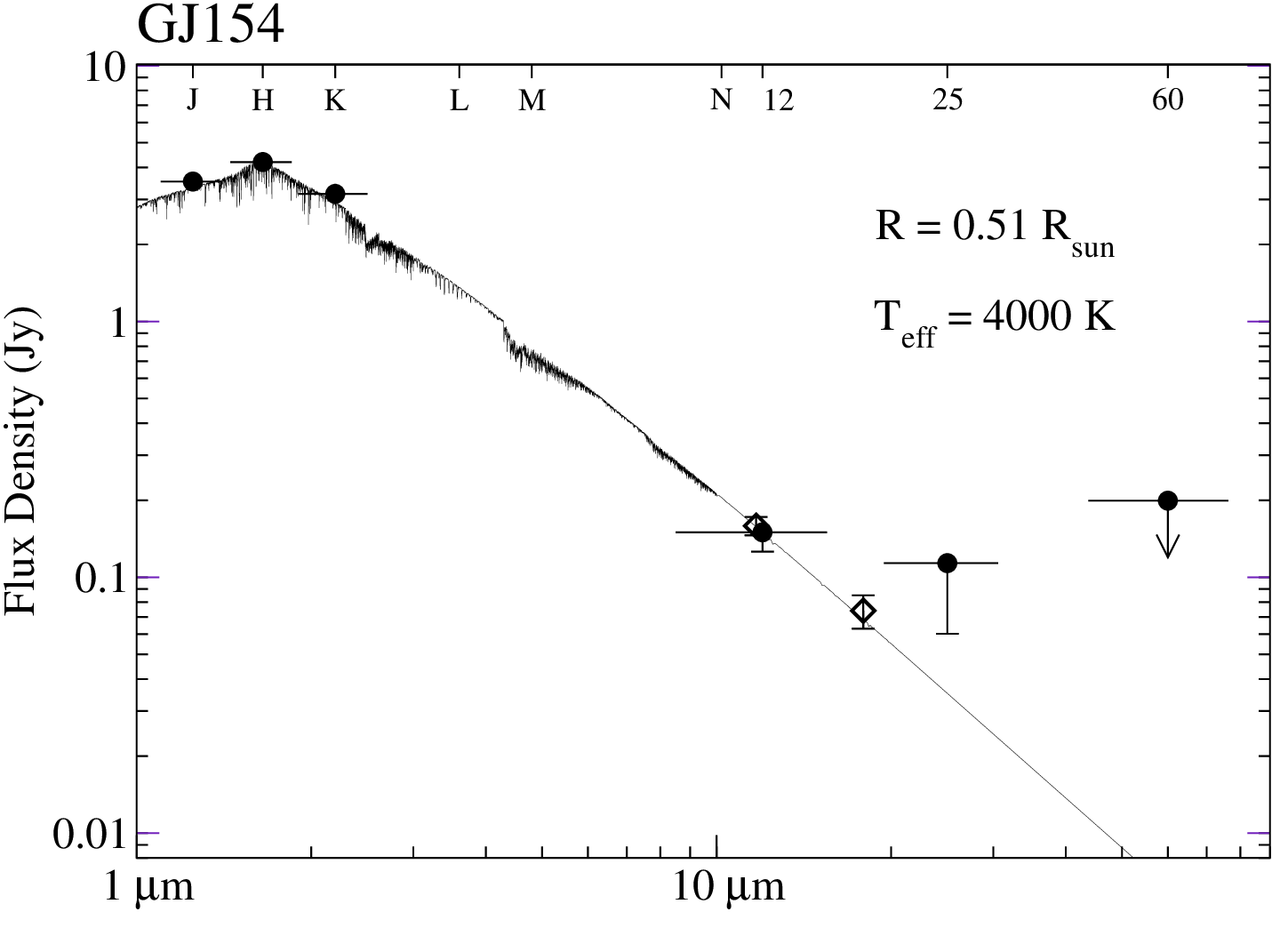}} &
\resizebox*{0.5\columnwidth}{!}{\includegraphics{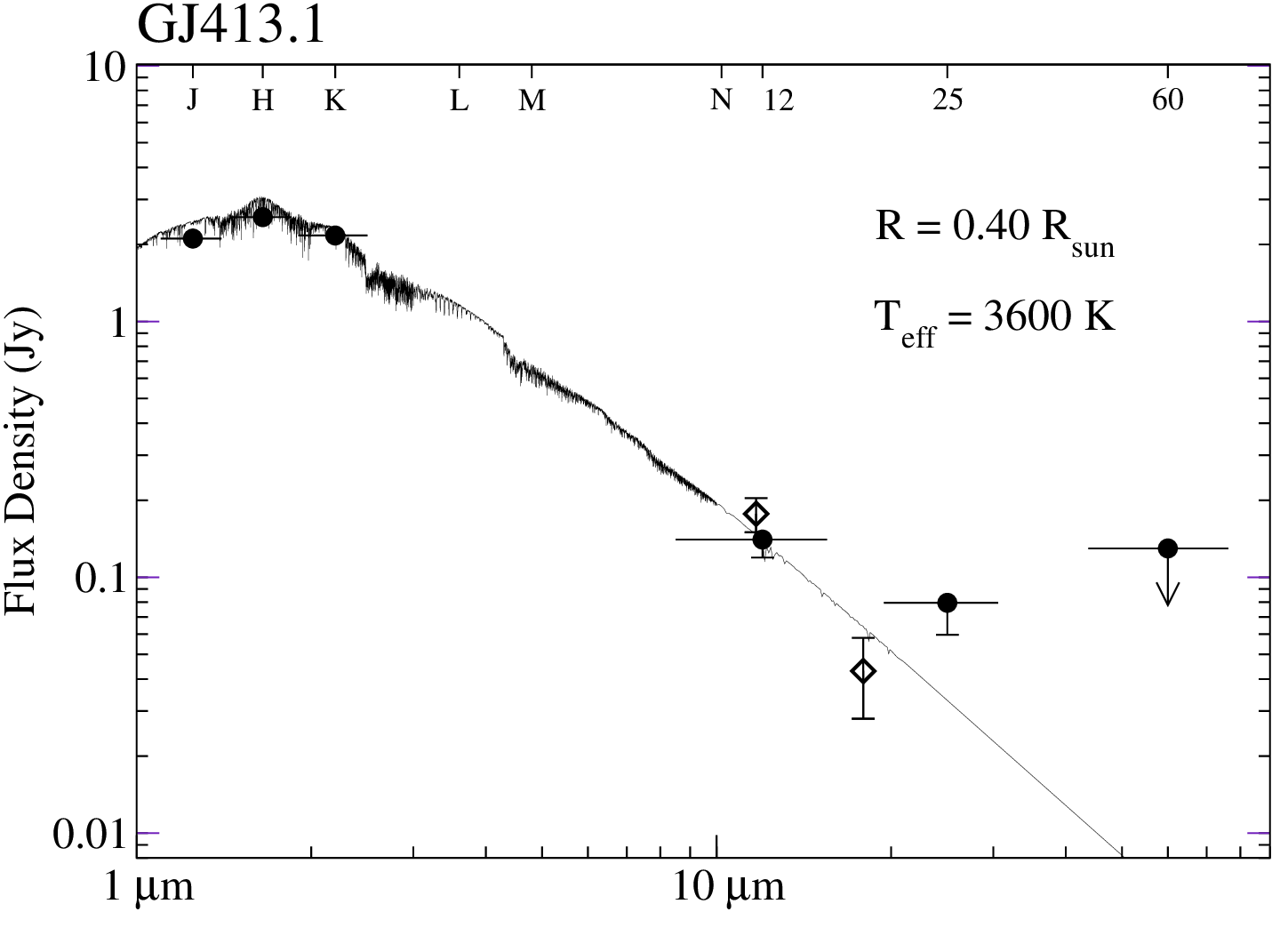}} \\
\resizebox*{0.5\columnwidth}{!}{\includegraphics{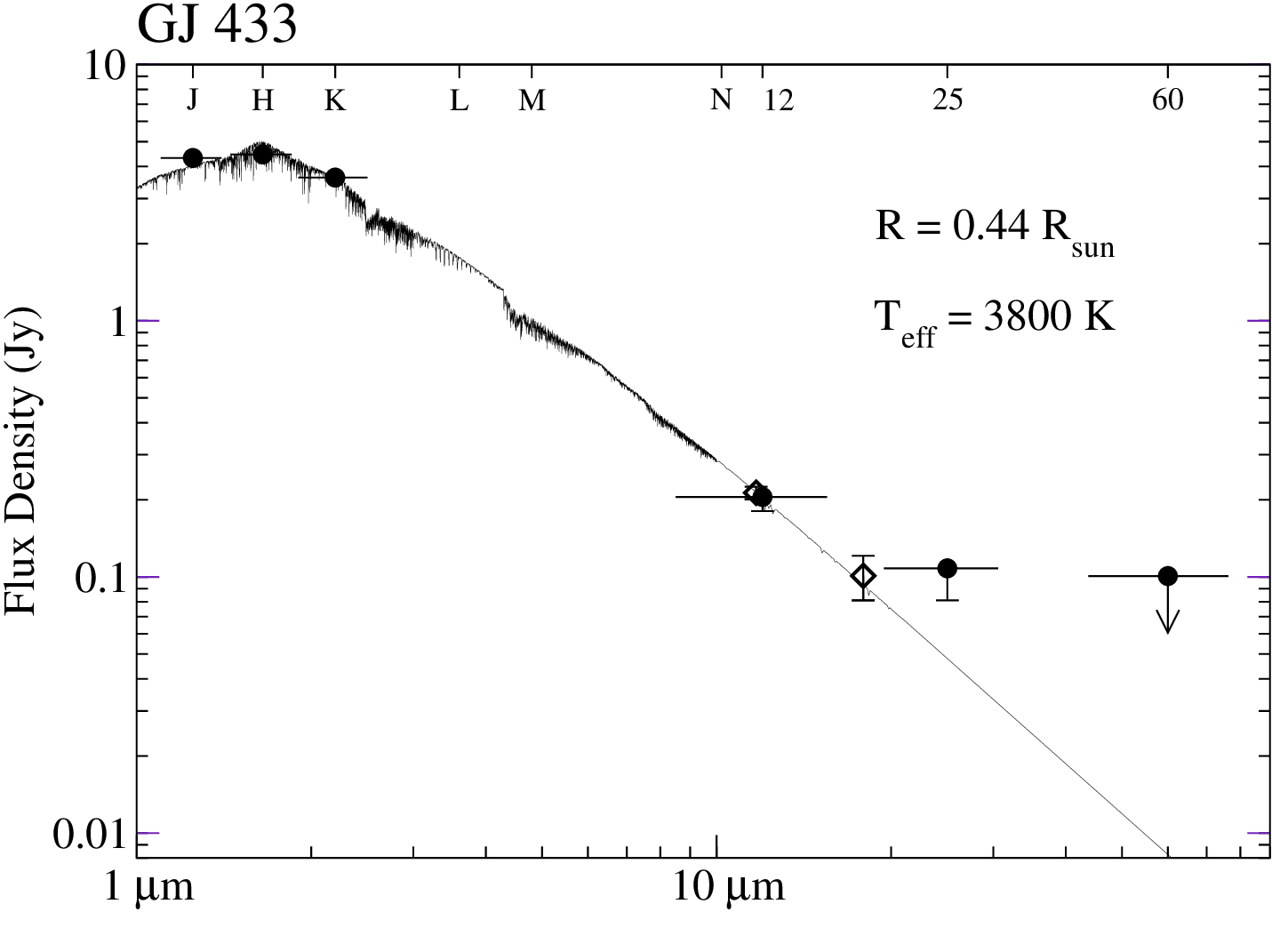}} &
\resizebox*{0.5\columnwidth}{!}{\includegraphics{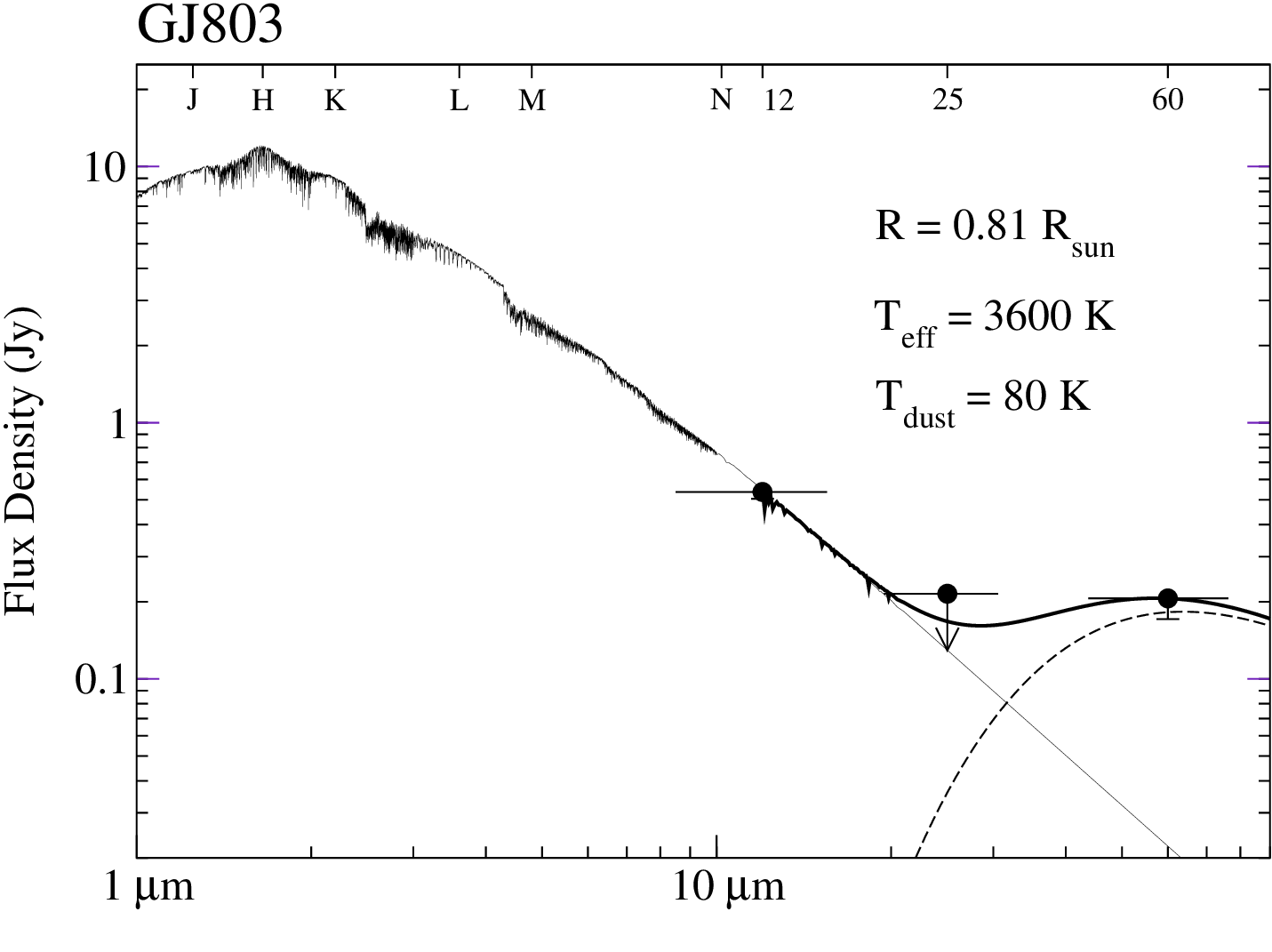}} \\
\end{tabular}\par}

\caption{Spectral energy distribution fits of M-type stars with tentative
IR excesses identified from this study by using the \emph{IRAS} FSC.
Solid circles are JHK and \emph{IRAS} data\emph{,} and diamonds indicate
our ground--based 11.7 and 17.9~\protect\( \mu \protect \)m fluxes.
Thin solid lines are synthetic stellar spectra fit to visual and near
IR (\protect\( \lambda <2\, \mu \mathrm{m}\protect \)) photometry
(\protect\( [M/H]=0.0\protect \) and \protect\( \log g=5.0\protect \))
and a dotted line (only for GJ~803) indicates a dust component with
\protect\( T=80\, K\protect \) and \protect\( \mathrm{L}_{\mathrm{IR}}/\mathrm{L}_{*}=6.7\times 10^{-4}\protect \).
Wavelength and flux density scales are logarithmic. Horizontal bars 
across JHK and IRAS data points indicate passband widths.\label{SEDs}}
\end{figure}

Photometry was performed in a 16 pixel (\( 1.''3 \)) diameter synthetic
aperture on each image and the results are reported in Table~\ref{table}.
For an M--star of luminosity \( 0.1\, \mathrm{L}_{\odot } \), blackbody-like grains
at a thermal equilibrium temperature of 200~K will sit 0.6~AU from
the star. Therefore, at a distance of 10~pc, any 12 or 18~$\mu$m
excess should appear \( <0.^{''}1 \) in size, or spatially unresolved.
It is clear from Figure~\ref{SEDs} that the apparent 25~$\mu$m
\emph{IRAS} excesses of GJ~154, 413.1, and 433 are not real. We interpret
this discord as follows: 

For the faint stars under consideration whose real fluxes are near
detection threshold, a downward noise fluctuation could place the
25~$\mu$m fluxes below the \emph{IRAS} detection threshold;
thus none displays a significant 25~$\mu$m flux deficit (negative
\( r \)).  Occasional large upward noise fluctuations could boost
25~$\mu$m fluxes so that they would be classified as IR excess
stars (positive \( r \), "Detection threshold bias" or "Malmquist bias"). The final configuration 
thus resembles our Figure~1, with some excess stars but with no significant deficit
star. In fact, the \emph{IRAS} 25~$\mu$m S/N ratios of all
of our three false IR excess stars are \( \sim 4 \) which is the
IRAS threshold value. 

\section{Statistical significance of IR excess}

\begin{figure}
{\centering \begin{tabular}{cc}
\subfigure[GJ 816]{\resizebox*{0.5\columnwidth}{!}{\includegraphics{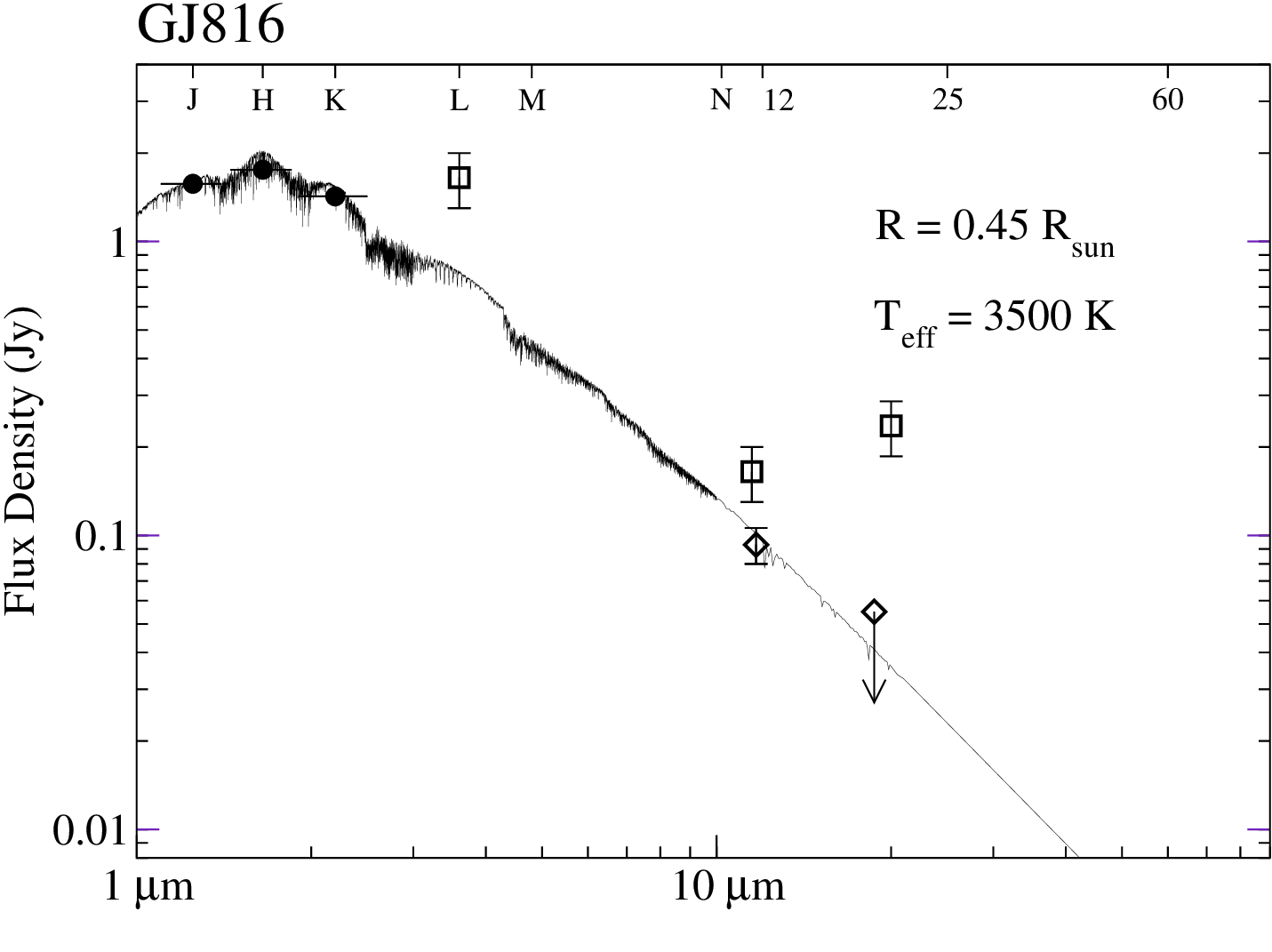}}} &
\subfigure[HD 2381]{\resizebox*{0.5\columnwidth}{!}{\includegraphics{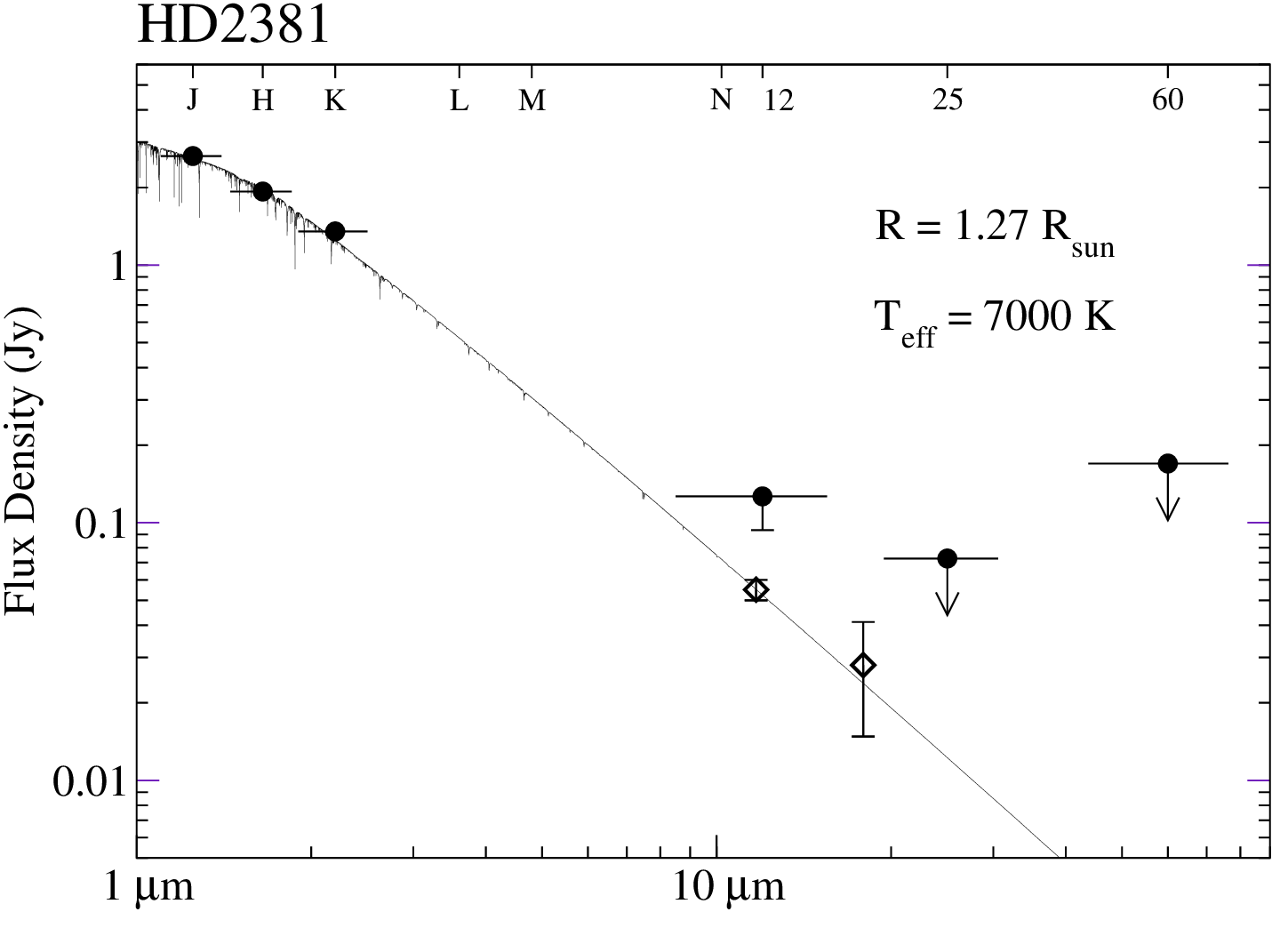}}} \\
\end{tabular}\par}

\caption{Spectral energy distribution fits of Vega--like stars from the literature.
Symbols have the same meaning as in Figure~\ref{SEDs}. For GJ~816,
open squares show \emph{ISO} fluxes (there is no \emph{IRAS} data)
and open diamonds are LWS measurements with the Keck~I
telescope. \label{FAs}}
\end{figure}

Recently, \citet{FA1,FA2} suggested that certain stars possess excess
emission as measured by IRAS or ISO. We checked IR excesses at GJ~816 and HD~2381
with 11.7 and 17.9~$\mu$m (18.7~$\mu$m for GJ~816)
Keck photometry. Apparent excesses for both stars turned
out to be false positives (Figure~\ref{FAs}). GJ~816 is not an
\emph{IRAS} FSC source and \citet{FA1} used \emph{Infrared} \emph{Space
Observatory (ISO)} data. An incorrect \emph{ISO} flux calibration
(for GJ~816) and Malmquist bias (for HD~2381) similar to our three
false IR--excess stars may be responsible for these apparent excesses.

An occasional large upward noise fluctuation (e.g., \( 2\sigma \approx 2\% \)
probability) does not significantly influence stars with high signal-to-noise
ratios; however, it can significantly affect stars with low signal-to-noise
data. For our initial 152 \emph{IRAS} sources, we expect \( \sim 3 \)
to have flux overestimates \( \ge 2\sigma  \), in apparent agreement
with what we have found. Based on this fact, some suggested Vega--like
stars --- generally identified through huge surveys often encompassing
thousands of input stars --- could also be non--IR--excess stars.
Thus, we suggest the following criteria for {\em bona--fide} Vega--like
stars; (1)~high S/N not subject to a Malmquist bias, (2)~low S/N
detections at 2 or more wavelengths, or (3)~ground/space--based confirmation
(e.g., Silverstone 2000 and this study) with higher sensitivity and better spatial
resolution than \emph{IRAS}.

\section{Summary and Discussion \label{S&D}}

We have performed a search for IR excess emission among M--type stars by correlating
the \emph{IRAS Faint Source Catalog} with \emph{Hipparcos} selected
late--type stars. Besides the previously known Vega--like star (GJ~803),
three tentative excess stars were identified, but these excesses turned
out to be false based on our ground--based mid--IR photomtery. Detection
threshold bias (Malmquist bias) is thought to be responsible for these bogus \emph{IRAS}
IR excesses. Two other stars (GJ~816 and HD~2381), suggested to be Vega--like in
the literature, are also shown to be non--IR--excess stars. In future
studies, one should be aware that some Vega--like stars reported in
the literature with low S/N ratios may be non--IR--excess stars as
well. This is likely to be the case for most stars listed by \citet{FA2}.

GJ~803 and Hen~3--600 show strong 60~$\mu$m excesses; they
are the only unambiguously identified M--type dwarf stars with IR excesses. This
could be due to the extreme youth of GJ~803 \citep[12~Myr, ][]{bPic2} and
Hen~3--600 \citep[$8-10$~Myr, ][]{TWA}. \citet{AAS_Pasadena}
have found two very young (\( \sim 12 \)~Myrs) late--type stars (HIP~23309,
M0 and HIP~29964, K6) co--moving with \( \beta  \)~Pictoris. Even
if one assumed that these two \emph{Hipparcos} stars have the same
fractional IR luminosity as \( \beta  \)~Pictoris (\( L_{IR}/L_{star}\sim 10^{-3} \)),
their corresponding 60~$\mu$m fluxes (\( <80 \)~mJy and \( <40 \)~mJy,
respectively) are below the \emph{IRAS} detection threshold. This
is true for late--type stars with $\beta$~Pic-like excess in nearby young stellar groups,
i.e, TWA. These stars would be excellent targets for future IR excess
surveys by \emph{SOFIA} or \emph{SIRTF}.

\acknowledgements{We are grateful to Mr. Michael Schwartz for assistance obtaining
JHK data for GJ~413.1 and GJ~433 and Dr. M. Jura for helpful discussion
and assistance with a Keck observation. We also thank an anonymous referee for 
suggestions that clarified the paper. This research was supported
by the UCLA Center for Astrobiology and by NASA. We have used the
SIMBAD/Vizier database.}

\onecolumn

\def\mlc{\multicolumn}
\begin{landscape}
\begin{table}

\caption{M--type IR--excess candidates \label{table}}

\scriptsize
{\centering \begin{tabular}{ccc c c c r@{$\pm$}l r@{$\pm$}l r@{$\pm$}l c r@{$\pm$}l r@{$\pm$}l  c c c}
\hline 
GJ& Sp.& dist & \multicolumn{3}{c}{near IR data (mag)$^*$} & 
                \multicolumn{6}{c}{IRAS flux (mJy) }   & 
                $r$ & 
                \multicolumn{4}{c}{Keck flux (mJy) }   &
                \multicolumn{2}{c}{Prediction (mJy)$^{**}$}   &
                excess? \\
\cline{4-6} \cline{7-12} \cline{14-17} \cline{18-19}
& Type & (pc) & {J} & {H} & {K} &
                \multicolumn{2}{c}{12~$\mu$m} & 
                \multicolumn{2}{c}{25~$\mu$m} & 
                \multicolumn{2}{c}{60~$\mu$m} & 
                 value & 
                \multicolumn{2}{c}{11.7~$\mu$m} & 
                \multicolumn{2}{c}{17.9~$\mu$m} &
                11.7~$\mu$m & 17.9~$\mu$m & \\
\hline

154    & M0 &14.6& 6.67(3)& 6.03(5)& 5.85(5)& 150&24 & 114&53 &\multicolumn{2}{c}{$<199$} & 2.18& 159&13&  74&11  & 157 & 68 & NO\\
413.1  & M2 &10.7& 7.23(2)& 6.55(2)& 6.23(2)& 141&21 &  80&20 &\multicolumn{2}{c}{$<130$} & 3.38& 177&27&  43&20  & 147 & 64 & NO\\
433    &M1.5& 9.0& 6.46(2)& 5.95(2)& 5.67(2)& 205&25 & 108&27 &\multicolumn{2}{c}{$<101$} & 2.28& 213&11& 101&20  & 214 & 93 & NO\\
803    & M0 & 9.9&   ---    &   ---    &   ---    & 537&32 &\multicolumn{2}{c}{$<215$} & 273&46 & 
       7.60\( ^{\dag } \)& \multicolumn{2}{c}{---}& \multicolumn{2}{c}{---}& 574 & 257 & YES\\
816    & M3 &13.8& 7.55(1)& 6.96(2)& 6.69(2)& \multicolumn{2}{c}{---}&
                                                    \multicolumn{2}{c}{---}&
                                                    \multicolumn{2}{c}{---}&
                                                    {---}& 93&13& \multicolumn{2}{c}{55\( ^{\flat } \)}& 101 & 45 & NO\\
HD 2381& F2V&74.2& 6.99(1)& 6.86(4)& 6.74(1)& 127&33 & \multicolumn{2}{c}{$<73$} & \multicolumn{2}{c}{$<170$} & 
                                                    2.30\( ^{\ddag } \)& 55&5 & 28&14&  55 & 24 & NO\\
\hline
\multicolumn{20}{l}{$^*$ near IR data for GJ 154 from \citet{Alonso}, for GJ 413.1 and GJ 433 from our IRTF measurements, and
                         for GJ 816 and HD 2381 from 2MASS database.} \\
\multicolumn{20}{l}{ \( ^{**} \) expected photospheric flux.}\\
\multicolumn{20}{l}{ \( ^{\dag } \) 60~$\mu$m spec. excess. 25~$\mu$m value is upper limit.}\\
\multicolumn{20}{l}{\( ^{\ddag } \) 12~$\mu$m spec. excess. 25~$\mu$m value is upper limit.}\\
\multicolumn{20}{l}{\( ^{\flat } \) this is 18.7~$\mu$m flux upper limit not 17.9~$\mu$m.}\\
\end{tabular}\par}
\end{table}
\end{landscape}

\end{document}